\documentclass{ws-procs975x65}
\usepackage[absolute]{textpos}
\textblockorigin{1.14\textwidth}{3.4cm}
\begin{document}

\title{\uppercase{Asymptotic Safety and Black Hole Thermodynamics}%
}

\author{D. BECKER and M. REUTER}

\address{Institute of Physics, University of Mainz,\\
D-55128 Mainz, Germany%
}
\vspace{-10pt}
\begin{abstract}
We present recent results on the non-perturbative renormalization group flow of Quantum Einstein Gravity (QEG) on spacetime manifolds with boundaries. As an application, novel quantum gravity corrections to the thermodynamics of black holes are discussed.\footnote{Talk given by M.R. at the Thirteenth Marcel Grossmann Meeting
, Stockholm, 2012.}
\end{abstract}


\bodymatter\bigskip

\begin{textblock}{1}(0,0)
\textblocklabel{block two}
MZ-TH/12-57
\end{textblock}

The Asymptotic Safety approach to quantum gravity\cite{Weinberg1979, Reuter1998b,Niedermaier2006} is based upon the idea that a fundamental quantum field theory of gravity can be constructed by taking the crucial UV limit at a suitable fixed point in  `theory space', the space of `all' action functionals.\cite{Reuter2007,Nink2012} 
This space is spanned by an infinite number of field monomials which form a basis and  respect the symmetries imposed on the action. 
A renormalization group (RG) trajectory on theory space, parametrized by the scale $k$, is a solution to a functional renormalization group equation (FRGE) and can be described by a set of scale dependent couplings, the coordinates in theory space, associated to the basis elements. 
In practice, only truncated theory spaces are considered which retain a  reduced (finite) number of such basis elements, for either the metric\cite{Reuter1998b,Reuter2007,Nink2012} or tetrad\cite{Tetrad} field contents. 
By now, a UV fixed point ---  along with a suitable trajectory that emerges from this fixed point in the UV ($k\rightarrow\infty$) and resembles General Relativity in the IR ($k \ll m_{\text{Pl}}$) --- was found for various kinds of truncations. Taken together they constitute a significant amount of evidence for the existence  of a UV fixed point in the full, un-truncated theory space, indicating the non-perturbative renormalizability of quantum gravity. 

Whereas so far investigations focused on spacetime manifolds without boundaries and thus only on the running couplings related to `bulk' invariants, in Ref. \refcite{Becker2012} we studied for the first time spacetimes exhibiting a non-empty boundary and scale dependent coefficients of associated surface invariants.
This not only allows to test the stability of the UV fixed point under a special change of topology but also makes a large class of physically interesting spacetimes accessible now, black holes, in particular.

We choose a bi-metric truncation\cite{Manrique2010} in the gravitational sector, coupled to a set of $N$ scalar matter fields $A_j$, and include a total of 17 running couplings, related to both bulk and boundary terms.
The bi-metric nature of the ansatz is essential to separate the effects due to  the (fixed, but arbitrary) background metric $\bar{g}_{\mu\nu}$ from those of the dynamical metric $g_{\mu\nu}\equiv\bar{g}_{\mu\nu}+\bar{h}_{\mu\nu}$, where $\bar{h}_{\mu\nu}$ is the (not necessarily small) fluctuation field. 
The background or `level-(0)' part of the effective average action $\Gamma_k$ is of zeroth order in $\bar{h}_{\mu\nu}$ and  contains the usual Einstein-Hilbert action, along with a Gibbons-Hawking like surface term and a non-minimally coupled scalar sector: \pagebreak
\begin{align}
 \Gamma_k^{\text{B}}[A;\bar{g}]
&=  \int_{\mathcal{M}}\text{d}^d x \sqrt{\bar{g}}\, \left\{\frac{1}{16\pi G_k^{(0)}} \Big(-\bar{R} + 2\Lambda^{(0)}_k\Big) +\frac{1}{2}\,\bar{g}^{\mu\nu} \partial_{\mu}A\partial_{\nu} A +\frac{1}{2} \xi_k^{(0)} \bar{R} A^2 + V_k^{(0)}\right\} \nonumber\\
 &\quad -  \int_{\partial\mathcal{M}}\!\!\!\!\text{d}^{d-1} x \sqrt{\bar{H}}\, \left\{ \frac{1}{16\pi G_k^{(0,\partial)}}\left(2\bar{K} - 2\Lambda^{(0,\partial)}_k\right) - \xi_k^{(0,\partial)}\bar{K} A^2\right\} \label{eqn:01_01}
\end{align}
Here, $\bar{K}=\bar{D}_{\mu}n^{\mu}$ is the trace of the extrinsic curvature in terms of the normal vector $n^{\mu}$. For the `level-(1)' terms, by definition linear in $\bar{h}_{\mu\nu}$, we make the ansatz:
\begin{align}
&\Gamma^{\text{lin}}_k[\bar{h},A;\bar{g}]=
 \frac{1}{16 \pi G_k^{(1)}}\int_{\mathcal{M}}\text{d}^dx\sqrt{\bar{g}}\, \, \mathcal{E}_k^{\mu\nu}[\bar{g},A]\, \,\bar{h}_{\mu\nu} \label{eqn:01_02}
 \\
&\qquad +\int_{\partial \mathcal{M}}\!\!\!\!\!\!\!\text{d}^{d-1}\! x\sqrt{\bar{H}}\left\{ \frac{1}{16\pi } \left(\tfrac{1}{G_k^{(1)}}-\tfrac{1}{G_k^{(1,\partial)}}\right)- \frac{1}{2}\left(\xi_k^{(1,\text{II})}-\xi_k^{(1,\partial)}\right)A^2    \right\} n^{\lambda}\partial_{\lambda}\bar{h}^{\mu}_{\phantom{\mu}\mu}\,. \nonumber
\end{align}
In the bulk contribution of equation \eqref{eqn:01_02} we employed the convenient abbreviation $
 \mathcal{E}_k^{\mu\nu}[\bar{g},A]\equiv{\bar{G}}^{\mu\nu}-\frac{1}{2}\, E_k \,\bar{g}^{\mu\nu} \bar{R} + \Lambda^{(1)}_k\,\bar{g}^{\mu\nu} - 8\pi G_k^{(1)}{\cal T}_k^{\mu\nu}[A;\bar{g}]
$.
Hereby $\bar{G}^{\mu\nu}=\bar{R}^{\mu\nu}-\frac{1}{2}\bar{g}^{\mu\nu}\bar{R}$  is the usual Einstein tensor of the background metric, and ${\cal T}_k^{\mu\nu}$ is the energy momentum tensor of the matter fields, which contains further running couplings.

Computing the beta-functions in the induced gravity approximation (large $N$ limit) a first main result is found: There does indeed exist a non-Gaussian fixed point in the gravitational sector, supporting the Asymptotic Safety conjecture for QEG also in spacetimes with non-empty boundaries. While this result is remarkable in its own right, the form of the $\Gamma_k$ ansatz also sheds light on the different r\^oles played by the various field monomials in \eqref{eqn:01_01} and  \eqref{eqn:01_02} which, classically, would all appear with the same coefficient, {\it the} Newton constant $G_{\rm N}$, but here enjoy a different RG evolution. This leads to a whole zoo of  running couplings $(G_k^{(0)},\, G_k^{(0,\partial)},\, G_k^{(1)},\, G_k^{(1,\partial)},\,\cdots)$ that needs to be understood.

Let us consider a {\it self-consistent background}, that is, a solution to the effective field equations for vanishing fluctuations:
$
\big( \delta \Gamma_k \slash \delta \bar{h}\big)[\bar{h}=0; \bar{g}_k^{\text{self-con}}]=0\, .
$
The effective interactions entering $\delta \Gamma_k \slash \delta \bar{h}$ are completely determined by the level-(1) couplings of $\Gamma_k^{\text{lin}}$, such as $G_k^{(1)}$ and $G_k^{(1,\partial)}$. 
On the other hand, the level-(0) couplings of $ \Gamma_k^{\text{B}}$, like $G_k^{(0)}$, $G_k^{(0,\partial)},\cdots$, appear in the partition function
$ \mathbb{Z}_k = \exp\big( - \Gamma_k [\bar{h}=0 ;\bar{g}_k^{\text{self-con}} ] \big)$,
which determines the effective spacetime structure. Another key result related to the different Newton type couplings is the following natural candidate for a running ADM mass of an asymptotically flat spacetime:
\begin{align}
 M_k\equiv -\frac{1}{8\pi G_k^{(0,\partial)}} \oint_{S^2_{\infty}} \!\! \big(K -K_0 \big) \, . \label{eqn:01_06}
\end{align}
It depends on the surface coupling $G_k^{(0,\partial)}$.
The motivation for the definition \eqref{eqn:01_06} is that, in terms of this $M_k$, all relations of semi-classical black hole thermodynamics keep their familiar form, but with $M_k$ replacing the classical mass. Hence the RG running of $G_k^{(0,\partial)}$ is responsible for the quantum gravity corrections to the semi-classical results. Remarkably, $G_k^{(0,\partial)}$ shows  exactly the {\it opposite} scaling behavior than the bulk coupling $G_k^{(0)}$.\cite{Reuter1998b} Starting with a positive IR value at $k=0$, we find that $M_k \propto 1\slash G_k^{(0,\partial)}$ decreases towards the UV and vanishes near  the Planck scale $m_{\text{Pl}}$. 
Since $M_k$ is inversely proportional to $G_k^{(0,\partial)}$, the {\it increasing} value of the latter at large $k$ values does not contradict the idea of gravitational antiscreening\cite{Reuter1998b}, the bulk Newton constant becoming smaller in the UV. In fact, the ADM mass approaching zero near the Planck scale nicely fits with the picture that the entire mass originates from a kind of  `gravitational dressing'.

For a Schwarzschild black hole the running ADM mass is determined by a $k$-independent constant of integration, $R_{\text{s}}$, with dimension of a length:
$  M^{\text{Schw.}}_k=R_{\text{s}} \slash (2 G_k^{(0,\partial)})$.
The main equations of semi-classical black hole thermodynamics then remain unaltered up to the substitution $M\rightarrow M_k$. The entropy, for instance, is
\begin{align}
 S=A \slash (4\,G_k^{(0,\partial)})= (4\pi R_{\text{s}})^2 \slash (4\,G_k^{(0,\partial)}) \, . 
\end{align}
It tends to zero while $k$ approaches $m_{\text{Pl}}$. At this scale the specific heat capacity turns positive indicating a possible thermodynamical stabilization.
Applications include the final state of Hawking evaporation, the information paradox, or the structure of possible cold remnants. These results also raise  questions concerning their relation  to  older ones obtained by an `RG improvement' of classical black holes\cite{Bonanno}. For further discussions we must refer to Ref. \refcite{Becker2012}.

\vspace{6pt}
\noindent {\bf Summary.}
Bi-metric actions, like the one we considered here, are not only mandatory in order to implement background independence,\cite{Manrique2010} but often  also crucial for obtaining a proper conceptual understanding. 
The degeneracy among the different classical r\^oles played by Newton's constant, for example, gets lifted at finite scales $k$, and a large variety of inequivalent Newton type couplings emerges: $G_k^{(0)},\, G_k^{(0,\partial)},\, G_k^{(1)},\, G_k^{(1,\partial)},\,\cdots$. In the future, we shall have to learn their interpretation and understand their scale dependencies. 
Furthermore, it turned out that the inclusion of running surface terms is unavoidable for a proper analysis of asymptotically flat self-consistent backgrounds, for instance. They are particularly relevant to black hole thermodynamics, where they give rise to a running ADM mass $M_k\propto 1\slash G_k^{(0,\partial)}\!\!,$ encoding the leading quantum gravity corrections to the semi-classical theory.

\bibliographystyle{ws-procs975x65}

\begin{thebibliography}{1}

\bibitem{Weinberg1979}
S.~Weinberg, in: {\em General Relativity, an Einstein 
Centenary
  Survey} (CUP, 1979).

\bibitem{Reuter1998b}
M.~Reuter, {\em Phys.Rev.} {\bf D57},p.~971 (1998).

\bibitem{Niedermaier2006}
M.~Niedermaier and M.~Reuter, {\em Living Rev.Rel.} {\bf 
9}, p.~5 (2006).

\bibitem{Reuter2007}
For a recent review and a detailed list of references, see \\
M.~Reuter and F.~Saueressig, {\em New J.Phys.} {\bf 14}, p. 
055022 (2012), and
  arXiv:1202.2274.

\bibitem{Nink2012}
A.~Nink and M.~Reuter, arXiv: 1212.4325, and arXiv: 1208.0031.

\bibitem{Tetrad}
J.-E. Daum and M.~Reuter, {\em Phys.Lett.} {\bf B710},p.~215 
(2012); U.~Harst and
  M.~Reuter, {\it JHEP} {\bf 1205}, p.~5 (2012); P.~Dona 
and R.~Percacci, 
  arXiv:1209.3649.

\bibitem{Becker2012}
D.~Becker and M.~Reuter, {\em JHEP} {\bf 1207}, p.~172 
(2012).

\bibitem{Manrique2010}
E.~Manrique and M.~Reuter, {\em Annals Phys.} {\bf 325}, p.~785 (2010);
  E.~Manrique, M.~Reuter and F.~Saueressig, {\it Annals 
Phys.} {\bf 326}, p.
  440 (2011), p. 463 (2011).

\bibitem{Bonanno}
A.~Bonanno and M.~Reuter, {\em Phys. Rev.} {\bf D60}, p. 
084011 (1999), {\bf D62}, p. 043008 (2000), {\bf D73}, p. 083005 (2006); 
M.~Reuter and E.~Tuiran,
  {\it Phys. Rev.} {\bf D83}, p. 044041 (2011).

\end{thebibliography}

\end{document}